# Strategy to control biases in prior event rate ratio method, with application to palliative care in patients with advanced cancer

Xiangmei Ma [1], Grace Meijuan Yang [2,3,4], Qingyuan Zhuang [2,5], Yin Bun Cheung [1,4,6]

1. Centre for Quantitative Medicine, Duke-NUS Medical School, National University of Singapore, 8 College Road, Singapore 169857
2. Division of Supportive and Palliative Care, National Cancer Centre Singapore, 30 Hospital Boulevard, Singapore 168583.
3. Lien Centre for Palliative Care, Duke-NUS Medical School, National University of Singapore, 8 College Road, Singapore 169857
4. Programme in Health Services & Systems Research, Duke-NUS Medical School, National University of Singapore, 8 College Road, Singapore 169857
5. Data and Computational Science Core, National Cancer Centre Singapore, 30 Hospital Boulevard, Singapore 168583.
6. Duke-NUS AI + Medical Science Initiative, Duke-NUS Medical School, National University of Singapore, 8 College Road, Singapore 169857

\* Corresponding author:

Professor Yin Bun Cheung, Duke-NUS Medical School, 8 College Road, Singapore 169857

Email: yinbun.cheung@duke-nus.edu.sg

**Abstract**

**Objectives**: Prior event rate ratio (PERR) is a method shown to perform well in mitigating confounding in real-world evidence research but it depends on several model assumptions. We propose an analytic strategy to correct biases arising from violation of two model assumptions, namely, population homogeneity and event-independent treatment.

**Study Design and Setting**: We reformulate PERR estimation by embedding a treatment-by-period interaction term in an analytic model for recurrent event data, which is robust to bias arising from unobserved heterogeneity. Based on this model, we propose a set of methods to examine the presence of event-dependent treatment and to correct the resultant bias. We evaluate the proposed methods by simulation and apply it to a de-identified dataset on palliative care and emergency department visits in patients with advanced cancer.

**Results**:

Simulation results showed that the proposed method could mitigate the two sources of bias in PERR. In the palliative care study, analysis by the Cox model showed that patients who had started receiving palliative care had higher incidence of emergency department visits than their match controls (hazard ratio 3.31; 95% confidence interval 2.78 to 3.94). Using PERR without the proposed bias control strategy indicated a 19% reduction of the incidence (0.81; 0.64 to 1.02). However, there was evidence of event-dependent treatment. The proposed correction method showed no effect of palliative care on ED visits (1.00; 0.79 to 1.26).

**Conclusions**:

The proposed analytic strategy can control two sources of biases in the PERR approach. It enriches the armamentarium for real-world evidence research.

**Plain Language Summary**

The prior event rate ratio is a promising approach to control confounding in real-world evidence research but it depends on several model assumptions. We propose an analytic strategy to mitigate the biases arising from violation of two of the assumptions. It can facilitate appropriate use of the prior even rate ratio and broaden its scope of application.

*Word count*: 2876

# 1. Introduction

Real-world data plays an important role in the evaluation of the safety and effectiveness of treatments.[1,2] A major challenge in using real-world data is the threat of confounding. The prior event rate ratio (PERR) is an approach for controlling measured or unmeasured confounders.[3,4] The method has been applied to evaluate a variety of treatments in real-world settings.[5-9]

The core idea of the PERR is to partition the person-time of patients who were treated with the intervention concerned during an observation period into two sections: prior to or after the initiation of treatment. Each treated person is matched to at least one control patient who have not received the treatment by the end of the observation period. The treated person's date of treatment initiation is the index date for partitioning the matched control's person-time into prior and post periods. The Cox model is used to compare time-to-first-event between the two groups of patients separately in the prior and post period, generating two hazard ratios (HR), $HR_{prior}$ and $HR_{post}$, respectively. The PERR estimate of treatment effect is the ratio of the HRs:

$$PERR\ HR = HR_{post}/HR_{prior}.$$

Under several model assumptions such as magnitude of confounding being constant between prior and post periods, PERR HR cancels out the confounding and reveals the true treatment effect. The bootstrap method is used to obtain confidence intervals for the estimate.

The Cox model suffers a selection bias due to unobserved heterogeneity, also called omitted variables or "frailty".[10-12] The PERR inherits the Cox model's implicit assumption of homogeneity and therefore the bias.[13-15] It also assumes that the outcome events do not affect treatment rate.[14,16,17] The violation of this assumption is referred to as event-dependent treatment (EDT). If the outcome event increases treatment rate, event rate in the prior period among the treated persons would appear high and PERR HR would bias downward, or vice versa.[14,16,17] Currently there is no method for correction of bias due to EDT in PERR.

This article aims to (a) reformulate the PERR model to a form that is robust to the bias arising from unobserved heterogeneity, (b) propose a novel method to correct the bias due to EDT, and (c) evaluate the performance of the novel analytic strategy using simulation and apply it to real data on palliative care (PC).

There has been strong interest in the potential impact of PC on acute healthcare utilization.[18,19] There has been mixed evidence so far.[18,20] It has been hypothesized that differences between real-world and trial settings could lead to under-estimation of the effect of PC in randomized trials.[21] Real-world evidence research may play an important role in the evaluation. We will apply the proposed analytic strategy to a de-identified real-world dataset on PC uptake (exposure) and emergency department (ED) visit (outcome event) in patients with advanced cancer.

## 2. Materials and Methods

*2.1. Statistical Models*

We begin with a reformulation of the original PERR method to a new form that requires only one Cox model with a treatment-by-period interaction term. This paves the way for adoption of the Andersen-Gill (AG) model, an extension of the Cox model for analysis of recurrent event data, which is robust to unobserved heterogeneity.[22-24] Based on the AG model, we propose a method to detect EDT and correct the resultant bias.

2.1.1. Reformulation of the original PERR

In the original PERR, two Cox models are estimated separately using time-to-first-event data in the prior and post periods. Note that a person may have a "first" event in the prior period and another "first" event in the post period.

We propose an alternative formulation based on a single Cox model with a treatment-by-period interaction term:

$$\lambda_i(t) = \lambda_0(t) \exp(\beta_1 \times trt_i + \beta_2 \times post_i + \beta_3 \times trt_i \times post_i) \quad (1)$$

where $\lambda_0(t)$ is the unspecified baseline hazard, $trt_i=1$ if the i-th participant was ever treated during the observation period and $trt_i=0$ otherwise, $post_i=1$ for person-time in the post period and $post_i=0$ otherwise. Participants' end times of the prior period are the times at their "first" event in the prior period or the treatment/index times, whichever earlier. The start time of the post period is not reset to zero.[13,25] The coefficients are estimated by maximizing the partial likelihood (PL) using standard statistical packages:[11,24]

$$PL(\beta) = \prod_{j=1}^{n} \frac{\exp(\beta_1 \times trt_{[j]} + \beta_2 \times post_{[j]} + \beta_3 \times trt_{[j]} \times post_{[j]})}{\sum_{k \in R(t_{(j)})} \exp(\beta_1 \times trt_k + \beta_2 \times post_k + \beta_3 \times trt_k \times post_k)} \quad (2)$$

where n is the total number of events, [j] indexes the person who had a "first" event at time $t_{(j)}$, $R(t_{(j)})$ is the set of persons who are at risk of "first" event at time $t_{(j)}$, $\exp(\beta_1)$ and $\exp(\beta_1+\beta_3)$ are $HR_{prior}$ and $HR_{post}$, respectively, and $\exp(\beta_3)$ is simply the PERR HR. Robust standard error for clustered data (up to 2 "first" events per person) is used to obtain confidence intervals. To differentiate them from the original PERR, we call them $HR_{prior}^{Cox}$, $HR_{post}^{Cox}$ and $PERR_{Cox}$ HR, respectively.

2.1.2. Bias arising from unobserved heterogeneity

The hazard function and partial likelihood of the AG model have the same forms in equations (1) and (2), respectively. It is also fitted by standard statistical packages. Unlike the Cox model, a history of having experienced outcome events at $t_{(j-1)}$ or earlier does not lead to exclusion from $R(t_{(j)})$ or later risk sets in the AG model. Robust standard error for clustered data is used for inference. We refer to the estimation of equation (1) using AG model as $PERR_{AG}$ and the estimates of $\exp(\beta_1)$, $\exp(\beta_1 + \beta_3)$ and $\exp(\beta_3)$ as $HR_{prior}^{AG}$, $HR_{post}^{AG}$ and $PERR_{AG}$ HR, respectively. Previous studies have shown that Cox model analysis of time-to-first-event suffers a selection bias,[10-12] while the AG model is robust to it.[23,26] The AG model is the method of choice as there always is some unobserved heterogeneity [10] and its impact is not trivial.[12]

Yu et al. [15] and Lin and Henley [13] proposed alternative PERR methods based on stratified/pairwise Cox models to control the bias due to unobserved heterogeneity. We use $PERR_{AG}$ as it provides a framework that facilitates our proposal for detection of and correction for bias arising from EDT.

2.1.3. Detection of and correction for event-dependent treatment

Let $h_{trt}(t)$ be the underlying instantaneous rate (hazard) of starting treatment at time t and $h_{trt}^*(t)$ be its realization in the presence of EDT, where

$$h_{trt}^*(t) \begin{cases} \theta \times h_{trt}(t) & \text{for } t_{ij} < t \leq t_{ij} + \delta, \\ h_{trt}(t) & \text{for } t \leq t_{ij} \text{ and } t_{ij} + \delta < t, \end{cases} \quad (3)$$

$t_{ij}$ is the j-th outcome event of the i-th participant, $\delta$ ($\geq 0$) is a duration the outcome event influences treatment rate, $\theta$ ($\geq 0$) denotes the magnitude of event occurrence on treatment rate, and $1<\theta$ and $\theta<1$ represent outcome events increase or decrease treatment rate, respectively. Assuming $\delta$ is finite, equation (3) leads to an estimate of the impact of preceding events on treatment time (derivation shown in Online Supplementary Materials 1):

$$\Delta t_i = t_{trt,i}^* - t_{trt,i} = \begin{cases} (1-\theta)\delta & \text{if } t_{ij} + \theta\delta < t_{trt,i}, \\ (t_{trt,i} - t_{ij})\left(\frac{1}{\theta} - 1\right) & \text{if } t_{ij} < t_{trt,i} \leq t_{ij} + \theta\delta, \end{cases} \quad (4)$$

where $t_{trt,i}$ is the participant's underlying treatment time in the absence of EDT, $t_{trt,i}^*$ is the realized treatment time in the presence of EDT, and $\Delta t_i$ is the shift of the treatment time caused by EDT. For example, if $\theta = 0.5$, $\delta = 10$, $t_{ij} = 100$ and $t_{trt,i}=110$ (days), the i-th participant's treatment would be delayed by $(1-0.5)\times10=5$ days, i.e. $t^*_{trt,i} =115$.

The impact of $1 < \theta$ or $\theta<1$ is empirically manifested in elevation or reduction of event rate in a duration of length $\delta$ before the observed start date of treatment.[27] Let the prior period for participant i be $t \in (A_i, B_i]$, where $A_i$ is the start of the prior period, $B_i = t^*_{trt,i}$ for treated persons and it equals index time for controls. To quantify the impact of EDT, we partition the prior period into sub-periods $(A_i, B_i - \hat{\delta}]$ and $(B_i - \hat{\delta}, B_i]$, where $\hat{\delta}$ is an estimate of $\delta$. Then, fit the AG model using data from the prior period only:

$$\lambda_i(t) = \lambda_0(t) \exp(b_1 \times trt_i + b_2 \times gap_i + b_3 \times trt_i \times gap_i), \quad (5)$$

where $gap_i=1$ for the duration $(B_i - \hat{\delta}, B_i]$, $gap_i=0$ otherwise, and $\exp(\hat{b}_3) = \hat{\theta}$ is an estimate of $\theta$.

However, we probably do not know a priori what $\hat{\delta}$ is. To find a suitable value for $\hat{\delta}$, we propose initial exploratory analysis with M>1 gaps of equal length $\dot{\delta}$ in AG model:

$$\lambda_i(t) = \lambda_0(t) \exp\left(b_1 \times trt_i + \sum_1^M b_{2m} \times gap_{mi} + \sum_1^M b_{3m} \times trt_i \times gap_{mi}\right), \quad (6)$$

where $\dot{\delta}$ is an initial guess such that $M\dot{\delta}$ is likely larger than $\delta$ according to subject-matter knowledge. Then the partitioned sub-periods are $(A_i, B_i - M\dot{\delta}]$, $(B_i - M\dot{\delta}, B_i - (M-1)\dot{\delta}]$, …, $(B_i - \dot{\delta}, B_i]$. By evaluation of the estimates $\hat{b}_{3m}$ (m=1,2,…M) and their statistical significance we determine a suitable $\hat{\delta}$ and then estimate $b_3$ in equation (5) using the chosen $\hat{\delta}$. To be clear, the first of M sub-periods of length $\dot{\delta}$ is the earliest and the M-th is immediately before the post period. With M coefficients for the sub-periods, the chance of false positive finding is inflated. We declare presence of EDT only if the coefficient for the M-th sub-period, $\hat{b}_{3M}$, is significant at 5% level, and $\hat{\delta}$ is sum of the width of consecutive sub-periods (including the M-th) whose $\hat{b}_{3m}$ estimates are significant at 5% level.

If there is evidence of EDT, we propose to correct the bias by introducing event-dependence in the index times in the control group based on the estimates $\hat{\delta}$ and $\hat{\theta}$ and equation (4). The original and modified control group datasets are referred to as data(control) and data(control*), respectively. The treatment group dataset is referred to as data(treat). The PERR$_{AG}$ is applied to compare data(treat) versus data(control*) instead of data(control). The comparison is fair as the treatment time in data(treat) and index time in data(control*) are both subject to event-dependence.

To illustrate the case of $\hat{\theta} > 1$, suppose the analysis aforementioned has found $\hat{\theta} = 2$ and $\hat{\delta} = 10$ days. Let $A_i$, $B_i$ and $C_i$ represent the i-th control group participant's prior period start time, index time, and post period end time (days), respectively, and $(t_{i1}, t_{i2}, ...)$ be the event times, if any. Suppose the participant had two events preceding the index time and $(A_i, B_i, C_i; t_{i1}, t_{i2}) = (0, 100, 200; 50, 60)$. To generate the participant's data in data(control$^*$), we shift $B_i$ forward by $(1-\theta)\delta = (1-2) \times 10 = -10$ from 100 to 90 as per equation (4) since $60+2\times10<B_i$. Description and illustration of the algorithm to generate data(control$^*$) for $\hat{\theta} < 1$ are available in Online Supplementary Material 2.

*2.2. Simulation*

We evaluated the proposed methods by simulation. A set of simulation scenarios were generated to resemble the PC study in the next section in terms of sample size, follow-up time in the prior and post periods, levels and pattern (increasing over time) of event rate and treatment rate. Then we varied the data patterns/parameters for greater generalizability. One thousand replicates were used for each scenario. Mean estimates, coverage probability of 95% confidence interval (CP), and root-mean-square-error (RMSE) are presented. Details of the simulation scenarios/parameters and procedures are available in Online Supplementary Materials 3 (S3.1). Stata codes for the simulation are available from https://github.com/cheungyb/PERR-PC-ED.

*2.3. Palliative care study*

We applied the proposed strategy to a de-identified electronic health record dataset from a study of hospital-based specialist PC for patients with advanced cancer in Singapore.[28] In Singapore, stopping treatments with curative intent is not a pre-requisite for PC. The dataset included dates of initiation of PC and ED visits of 2706 patients with stage 4 cancer treated at the National Cancer Center Singapore in the period from December 2017 to August 2022, of whom 135 patients who had received PC prior to December 2017 were excluded. Among the 2571 remaining, 1409 initiated PC and 1162 did not receive PC during the study period. Using age (at 1 December 2017; in 5-year band), gender and type of cancer as matching variables and a 1:1 matching ratio, subject to the condition that a control person must be still under follow-up at the time the matched treated person took up PC, a PERR dataset with 1658 patients was generated, including 829 treated persons and their matched controls.

To uphold the time-constant confounding assumption, the PERR literature has recommended restricting study duration to mitigate the possibility of changes in magnitude of confounding over time.[4,8] In the PC study, the mean (maximum) follow-up time in the matched dataset was 673 (1734) days. We restricted the PERR analysis to up to 150 days before and up to 150 days after treatment/index time.

## 3. Results

*3.1. Simulation results*

Table 1 shows that the original PERR and $PERR_{Cox}$ gave practically identical results regardless of baseline event rate ($\beta_0$) and level of heterogeneity ($\sigma_\omega^2$). Furthermore, both the original PERR and $PERR_{Cox}$ were biased in the presence of unobserved heterogeneity ($\sigma_\omega^2 \neq 0$). In contrast, $PERR_{AG}$ was not biased by unobserved heterogeneity and it gave smaller RMSE than the other two methods.

Table 2 shows the performance of the proposed method for detection and estimation of EDT, with M=5 and $\dot{\delta}$=10 days, where the true EDT effect was $\theta$=1/4, 1/2, 2 or 4 and true duration was $\delta$=20 or 30 days. For scenarios with strong EDT effect ($\theta$=1/4 or 4), there was about 92% to 96% probability of correctly estimating the duration of EDT ($\hat{\delta} = \delta$). For weak EDT effect ($\theta$=1/2 or 2), the probability of correctly estimating the duration was lower, at about 55% to 82%. But the probability of over- or under-estimation of the duration by more than one $\dot{\delta}$ was small even for weak EDT effect, at about 4% to 29%. Although the proposed method assumes a single $\theta$ value over the entire duration $\delta$, the simulation results show that it worked quite well even when truly EDT impact changed over the duration (two $\theta$s, one for half of $\delta$). The method also worked well in the negative control scenario ($\delta$=0, $\theta$=1).

Table 3 shows the performance of correction for EDT using the estimated $\hat{\delta}$ and $\hat{\theta}$. The true treatment effect was 0.5, i.e. 50% reduction of event rate. In all scenarios with EDT, the crude estimates of $PERR_{AG}$ HR without correction for ETD showed the expected direction of bias. Coverage probability (CP) of their 95% confidence intervals was as low as 38.6%. The proposed correction method gave mean $PERR_{AG}$ HR close to the true treatment effect, CP much closer to 95% and smaller RMSE. For the scenarios with weak EDT effect ($\theta$=1/2 or 2) that the estimation of impact duration was less accurate (Table 2), the bias in the uncorrected $PERR_{AG}$ HR estimate was relatively small and the corrected estimates were still close to the true value.

Further simulation results that varied the scenarios/parameters are available in Online Supplementary Materials 3 (S3.2, Tables S1 to S11). The results are similar to the aforementioned.

*3.2. Palliative care study*

In the post-period, incidence of ED visits was 3.39 and 1.07 visits per person-year in the PC and non-PC groups, respectively (Table 4). Patients who had started receiving PC had higher incidence of ED visits than their matched controls according to the Cox model, with $HR_{post}^{Cox}$ = 3.31 (95% CI 2.78 to 3.94). But they were at higher risk in the prior period to begin with: Incidence of ED visits was 3.59 and 0.78 in the PC and non-PC groups in the prior period, respectively, and $HR_{prior}^{Cox}$ = 4.12 (95% CI 3.50 to 4.85). The $PERR_{Cox}$ HR was 3.31/4.12 = 0.81 (95% CI 0.64 to 1.02). The corresponding estimates from the AG model for all events are $HR_{prior}^{AG}$ = 4.63 (95% CI 3.92 to 5.47), $HR_{post}^{AG}$ = 3.18 (95% CI 2.69 to 3.77), and $PERR_{AG}$ HR = 3.18/4.63 = 0.69 (95% CI 0.55 to 0.86), which was about 15% smaller than the $PERR_{Cox}$ HR estimate.

However, using equation (6) for the AG model to evaluate event-dependent treatment found that ED rate was elevated in a 5-week period before treatment (Figure 1). Refitting equation (5) with a pre-treatment window of $\hat{\delta}$=5 weeks gave estimate $\hat{\theta}$=4.76. Applying the proposed correction for event-dependent treatment gave adjusted $PERR_{AG}$ HR = 1.00 (95% CI 0.79 to 1.26).

**4. Discussion**

The PERR method is a promising approach to provide real-world evidence.[3,4,8] However, it requires several model assumptions. In this work we dealt with two of them. This enriches the armamentarium for real-world evidence research.

The simulation demonstrated that the original PERR can be implemented as one Cox model with a treatment-by-period interaction term, and the confidence interval can be estimated analytically. This not only simplifies the procedure and saves computation time, but also facilitates extension to the AG model which is robust to the bias arising from unobserved heterogeneity. The PC study also demonstrated a non-trivial difference between the Cox and AG model-based implementation of PERR.

The simulation showed that EDT can generate substantial bias and the proposed correction can mitigate the bias in treatment effect estimate, improve the coverage probability of the confidence intervals, and reduce the RMSE. The PC study demonstrated the practical impact of the correction method, changing an interpretation of protective effect to null effect.

A limitation of the correction method is that it does not completely remove the bias and the coverage probability of the confidence interval may still be somewhat different from the target level. Other limitations include that the correction is heuristic without mathematical proof, and that there is no definite guideline on the number and width of sub-periods (M and $\hat{\delta}$) in using equation (6) for detection of EDT. This requires initial guesstimate according to clinical context.

PERR also requires an assumption that event rate in the prior period does not affect event rate in the post period,[14,16] which we have not dealt with. Currently there is no correction method when this assumption is violated. Our study of palliative care and ED visits has not taken this issue into account and therefore is inconclusive. They are areas for further research in PERR methodology and in palliative care.

**Ethics Approval**

SingHealth Centralised Institutional Review Board (CIRB) initially reviewed and approved the palliative care study (CIRB reference number: 2017/2799), and subsequently deemed that further review was not necessary as the study was categorised as a health service evaluation.

**Data sharing and data availability statement**

Stata codes for the simulation reported are available from https://github.com/cheungyb/PERR-PC-ED. Data from the palliative care study are available from the author GMY upon reasonable request.

**Funding**

The methodological work was supported by the National Medical Research Council, Singapore (MOH-001487).

**Declaration of Interests**

None.

**ORCID**

Xiangmei Ma https://orcid.org/0000-0001-6526-1226

Grace Meijuan Yang https://orcid.org/0000-0002-0915-6007

Qingyuan Zhuang https://orcid.org/0000-0001-8700-9968

Yin Bun Cheung https://orcid.org/0000-0003-0517-7625

Table 1. Simulation results on estimates of treatment effect obtained from original PERR, $PERR_{Cox}$ and $PERR_{AG}$, by log-baseline hazard ($\beta_0$) and level of unobserved heterogeneity ($\sigma^2_\omega$); treatment effect = $\exp(\beta_3) = 0.5$.

| $\beta_0$ | $\sigma^2_\omega$ | PERR | | | $PERR_{Cox}$ | | | $PERR_{AG}$ | | |
|---|---|---|---|---|---|---|---|---|---|---|
| | | Mean estimate | CP (%)[a] | RMSE | Mean estimate | CP (%) | RMSE | Mean estimate | CP (%) | RMSE |
| -6.5 | 0 | 0.521 | 93.5 | 0.051 | 0.521 | 93.5 | 0.051 | 0.511 | 95.3 | 0.040 |
| -6.5 | 0.5 | 0.557 | 76.0 | 0.074 | 0.558 | 75.4 | 0.074 | 0.510 | 93.2 | 0.041 |
| -6.5 | 1.0 | 0.585 | 51.7 | 0.096 | 0.585 | 49.1 | 0.097 | 0.509 | 94.3 | 0.045 |
| -7.5 | 0 | 0.518 | 93.8 | 0.073 | 0.518 | 93.6 | 0.073 | 0.514 | 94.1 | 0.067 |
| -7.5 | 0.5 | 0.537 | 92.3 | 0.072 | 0.538 | 91.8 | 0.072 | 0.509 | 95.7 | 0.059 |
| -7.5 | 1.0 | 0.562 | 82.8 | 0.087 | 0.562 | 81.9 | 0.088 | 0.511 | 93.0 | 0.064 |

[a] Based on 200 bootstrap replicates

Table 2. Simulation results on strategy to estimate event impact on observed treatment time in PERR$_{AG}$ estimation, by levels of $\delta$ and $\theta$; M = 5 and $\dot{\delta}$ = 10 days.

| $\delta$ | $\theta$ | $P(\hat{\delta} = \delta)$ | $P(|\hat{\delta} - \delta| \leq \dot{\delta})$ | Mean $\hat{\theta}|_{\hat{\delta}=\delta}$ | Mean $\hat{\theta}|_{|\hat{\delta}-\delta|\leq\dot{\delta}}$ | Mean $\hat{\theta}$ |
|---|---|---|---|---|---|---|
| 0 [a] | 1 [a] | 94.6 | 99.6 | NA | NA | NA |
| 20 | 1/4 | 93.2 | 99.1 | 0.245 | 0.252 | 0.254 |
| 20 | 1/2 | 63.7 | 82.0 | 0.455 | 0.464 | 0.480 |
| 20 | 2 | 81.8 | 95.9 | 2.017 | 2.006 | 1.944 |
| 20 | 4 | 95.5 | 99.3 | 3.565 | 3.536 | 3.512 |
| 30 | 1/4 | 91.5 | 99.6 | 0.242 | 0.250 | 0.251 |
| 30 | 1/2 | 55.0 | 71.3 | 0.449 | 0.454 | 0.480 |
| 30 | 2 | 67.8 | 88.5 | 2.038 | 2.022 | 1.965 |
| 30 | 4 | 95.6 | 100.0 | 3.515 | 3.498 | 3.498 |
| 30 [b] | 1/4, 1/2 [b] | 77.3 | 96.5 | 0.339 | 0.337 | 0.335 |
| 30 [c] | 4, 2 [c] | 59.9 | 99.8 | 2.877 | 2.927 | 2.928 |

[a] Negative control: events have no impact on treatment.
[b] $\theta$ = 1/4 for 15 days followed by $\theta$ = 1/2 for another 15 days.
[c] $\theta$ = 4 for 15 days followed by $\theta$ = 2 for another 15 days.

Table 3. Simulation results on estimation of treatment effect by PERR$_{AG}$ with and without correction for event-dependent treatment bias, by levels of $\delta$ and $\theta$; true treatment effect = $\exp(\beta_3) = 0.5$.

| $\delta$ | $\theta$ | Uncorrected | | | Corrected based on $(\hat{\delta}, \hat{\theta})$ | | |
|---|---|---|---|---|---|---|---|
| | | Mean estimate | CP (%) | RMSE | Mean estimate | CP (%) | RMSE |
| 0 [a] | 1 [a] | 0.510 | 94.8 | 0.050 | 0.509 | 94.9 | 0.049 |
| 20 | 1/4 | 0.575 | 72.7 | 0.092 | 0.484 | 93.5 | 0.051 |
| 20 | 1/2 | 0.548 | 86.0 | 0.072 | 0.504 | 94.8 | 0.050 |
| 20 | 2 | 0.468 | 90.4 | 0.053 | 0.504 | 94.8 | 0.049 |
| 20 | 4 | 0.433 | 66.1 | 0.079 | 0.504 | 93.8 | 0.054 |
| 30 | 1/4 | 0.626 | 38.6 | 0.141 | 0.474 | 91.4 | 0.058 |
| 30 | 1/2 | 0.573 | 71.4 | 0.092 | 0.505 | 88.0 | 0.066 |
| 30 | 2 | 0.454 | 82.5 | 0.063 | 0.497 | 94.3 | 0.051 |
| 30 | 4 | 0.416 | 52.0 | 0.094 | 0.493 | 94.2 | 0.053 |
| 30 [b] | 1/4, 1/2 [b] | 0.601 | 54.2 | 0.117 | 0.483 | 90.9 | 0.059 |
| 30 [c] | 4, 2 [c] | 0.435 | 70.0 | 0.077 | 0.502 | 93.9 | 0.052 |

[a] Negative control: events have no impact on treatment.
[b] $\theta = 1/4$ for 15 days followed by $\theta = 1/2$ for another 15 days.
[c] $\theta = 4$ for 15 days followed by $\theta = 2$ for another 15 days.

Table 4. Incidence of emergency department visits (visits/person-years) by palliative care (PC) groups and periods, Singapore, 2017 to 2022.

| Period | PC | Non-PC |
| --- | --- | --- |
| Post | 3.39 (441/130) | 1.07 (335/312) |
| Prior | 3.59 (1028/286) | 0.78 (222/286) |

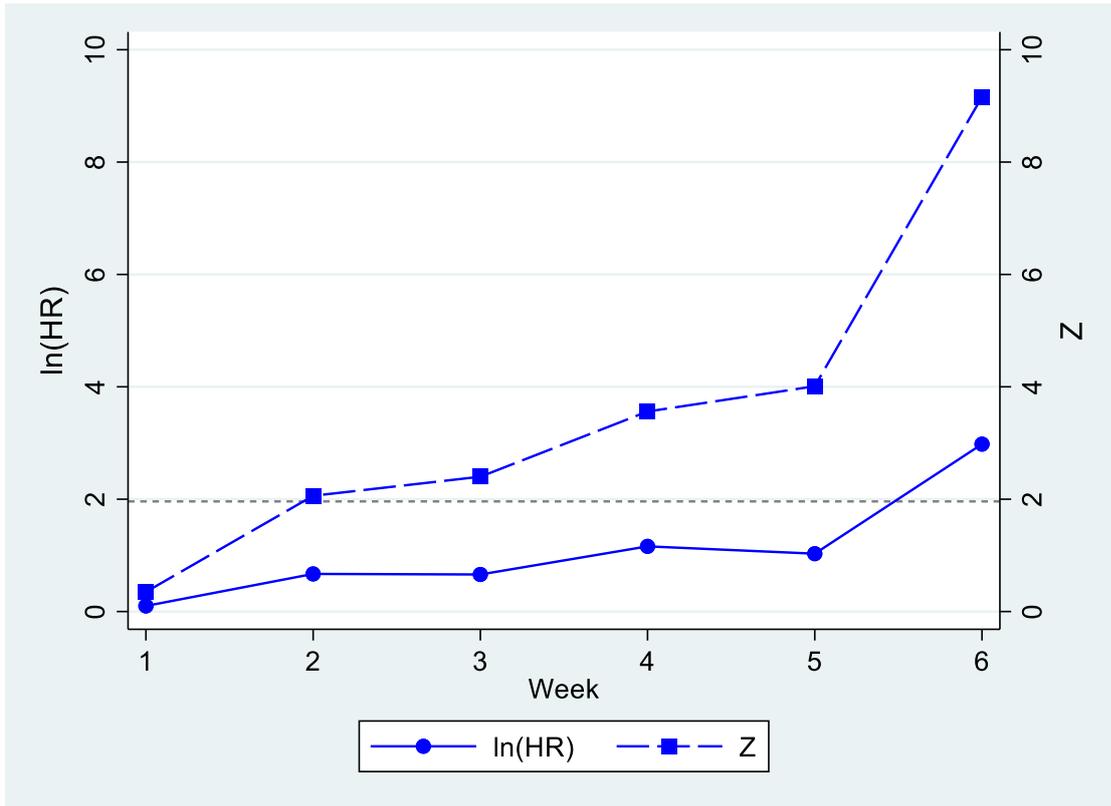

Figure 1. Evaluation of event-dependent treatment using equation (6), with M = 6 and $\dot{\delta}$ = 1 week; horizontal dotted line indicates Z-value 1.96.

**Online Supplementary Materials 1.** Derivation of the impact of event-dependent treatment on observed treatment time.

Let $t_{trt,i}$ be the i-th participant's underlying treatment time in the absence of event-dependent treatment (EDT), $t^*_{trt,i}$ be the realized treatment time in the presence of EDT, and $t_{ij}$ is the j-th outcome event of the i-th participant. Assumes that the instantaneous rate (hazard) of treatment is constant over $(t_{ij}, t_{trt,i})$ if events increase treatment rate:

$$h_{trt}(t \mid t_{ij} < t < t_{trt,i}) = h_{ij},$$

or constant over $(t_{ij}, t^*_{trt,i})$ if events reduce treatment rate:

$$h_{trt}(t \mid t_{ij} < t < t^*_{trt,i}) = h_{ij}.$$

If $t_{trt,i} < t_{ij}$, the event has no impact on treatment time.

If $t_{trt,i} > t_{ij}$,

$$\int_{t_{ij}}^{t^*_{trt,i}} \theta h_{ij}\, ds = \int_{t_{ij}}^{t_{trt,i}} h_{ij}\, ds,$$

which implies $\theta(t^*_{trt,i} - t_{ij}) = t_{trt,i} - t_{ij}$. That is $t^*_{trt,i} = \frac{t_{trt,i} - t_{ij}}{\theta} + t_{ij}$ if $t^*_{trt,i} \leq t_{ij} + \delta$.

If $t^*_{trt,i} > t_{ij} + \delta$, then

$$\int_{t_{ij}}^{t_{ij}+\delta} \theta h_{ij}\, ds + \int_{t_{ij}+\delta}^{t^*_{trt,i}} h_{ij}\, ds = \int_{t_{ij}}^{t_{trt,i}} h_{ij}\, ds,$$

which implies $\theta\delta + (t^*_{trt,i} - t_{ij} - \delta) = t_{trt,i} - t_{ij}$, that is $t^*_{trt,i} = t_{trt,i} - \delta(\theta - 1)$.

In conclusion,

$$\Delta t_i \triangleq t^*_{trt,i} - t_{trt,i} = \begin{cases} (1-\theta)\delta & \text{if } t_{ij} + \theta\delta < t_{trt,i}, \\ (t_{trt,i} - t_{ij})(\frac{1}{\theta} - 1) & \text{if } t_{ij} < t_{trt,i} \leq t_{ij} + \theta\delta, \end{cases}$$

where $\Delta t_i$ is the shift of the treatment time caused by EDT.

**Online Supplementary Materials 2.** Algorithm for generation of data(control$^*$) when event reduces treatment rate.

The procedure for correction for event-dependent treatment (EDT) when $\hat{\theta} < 1$ is not a simple opposite of the procedure for $\hat{\theta} > 1$. As can be seen in equation (4), the correction for EDT when $\hat{\theta} > 1$ shifts the index time of a control person forward but not preceding the previous event. However, application of the equation to correct for EDT when $\hat{\theta} < 1$ may or may not shift the index time to after the next event.

When $\hat{\theta} < 1$, for each $t_{ij} < B_i$, where $B_i$ is the index time of the i-th person, the index time is to be shifted backward by $\Delta t_i$ as indicated by equation (4), starting from $t_{i1}$. If the shifts lead to $B_i > C_i$, where $C_i$ is the end of the post period, the control person and the matched treated person are excluded from the analysis. To illustrate, suppose $\hat{\theta} = 0.5$, $\hat{\delta} = 10$ and $(A_i, B_i, C_i; t_{i1}, t_{i2}) = (0, 100, 200; 50, 102)$. To generate the participant's data in data(control$^*$), the index time is shifted backward by (1−0.5)×10=5, from 100 to 105 as $t_{i1} + \theta\delta < 100$, and then further shifted by (105−102)×(1/0.5−1)=3, from 105 to 108 as $t_{i2} < 105 < t_{i2} + \theta\delta$.

Note that for $\hat{\theta} > 1$ the shift in index time is made only for the immediate preceding event, whereas for $\hat{\theta} < 1$ we propose to shift the index time for each event preceding the index time. We experimented with making the shift only for the immediate preceding event or for all preceding event for $\hat{\theta} < 1$ and empirically we found that the latter performed substantially better in terms of bias and RMSE than the former. Hence the proposed method.

**Online Supplementary Materials 3.** Supplemental methods on simulation and supplemental simulation results.

## S3.1. Supplemental procedures

Generation of time-to-event and time-to-treatment data

Let $\tau_i$ be the censoring time of the i-th participant.

Time to treatment of the i-th participant, $t_{trt,i}$, is generated from the hazard function of a Weibull distribution:

$$h_{trt}(t) = k_1 t^{k_1-1} \exp(c_0 + \alpha_1 X_i + 0.5 Z_i + \epsilon_i), \quad (1)$$

where $X_i$ is an unobserved binary confounder with $Prob(X_i = 1) = 0.5$, $Z_i$ is an observed binary covariate with $Prob(Z_i = 1) = 0.5$, and $\epsilon_i$ is an independent random variation with mean 0 and variance 0.1. Define the time-varying treatment indicator to be

$$P_i(t) = 1 \ if\ t_{trt,i} < t \leq \tau_i; \text{otherwise } P_i(t) = 0.$$

Time to outcome events $\{t_{ij}: i = 1,2,\ldots,n; j = 1,2,\ldots,n_i\}$ till $\tau_i$ is generated from the hazard function of a Weibull distribution:

$$\lambda_i(t) = 0.5 k_2 t^{k_2-1} \exp(\beta_0 + \beta P_i(t) + \alpha_2 X_i + 0.5 Z_i + w_i), \quad (2)$$

where $w_i$ is the independent random variation with mean 0 and variance $\sigma_\omega^2$.

Data generation begins with generating observed follow-up time, $\tau_i \sim \text{uniform}(a, b)$. Then $t_{trt,i}$ and $t_{ij}$ are iteratively generated and updated by the inversion of cumulative distribution function method:

Step 1: Generate treatment time $t_{trt,i} \sim \left[\frac{-ln(\phi_i)}{h_{trt}(t)}\right]^{1/\kappa_1}$, where $\phi_i \sim \text{uniform}(0,1)\ \forall\ i$.

Step 2: Generate $t_{ij}$ for j=1, by $t_{ij} \sim \left[\frac{-ln(u_{ij1})}{\lambda_i(t)} + l^{\kappa_2}\right]^{1/\kappa_2}$, where $u_{ijk} \sim \text{uniform}(0,1)\ \forall\ i,j,k$ and $l=0$.

Step 3:

- If $t_{ij} < t_{trt,i}$ and $\theta \neq 1\ \&\ 0 < \delta$, (a) update $t_{trt,i}$ by $t^*_{trt,i} \sim \left[\frac{(t_{trt,i})^{\kappa_1} - (t_{ij})^{\kappa_1}}{\theta} + (t_{ij})^{\kappa_1}\right]^{1/\kappa_1}$; (b) if the updated $t^*_{trt,i}$ is larger than $t_{ij} + \delta$, update $t_{trt,i}$ by

  $t^*_{trt,i} \sim \left[(t_{trt,i})^{\kappa_1} + (1-\theta)\left((t_{ij} + \delta)^{\kappa_1} - (t_{ij})^{\kappa_1}\right)\right]^{1/\kappa_1}$.

- If $t_{ij-1} < t_{trt,i} \leq t_{ij}$, update $t_{ij}$ by $t_{ij} \sim \left[\frac{-ln(u_{ij2})}{\lambda_i(t)} + t_{trt,i}{}^{\kappa_2}\right]^{1/\kappa_2}$, where $t_{i0} = 0$ and $\lambda_i(t)$ has updated time-varying covariate $P_i(t) = 1$ as defined in equation (2).

Step 4: Reset $l = t_{ij}$, then reset $j = j + 1$ and $t_{trt,i} = t^*_{trt,i}$; repeat Step 2 to Step 3 till $\tau_i < t_{ij}$ and set $(t_{ij-1}, \tau_i]$ as the final spell of follow-up, which is censored.

Finally, determine that the participant is eligible for inclusion as a treated person if $t_{trt,i} < \tau_i$; otherwise the participant is eligible as a control person.

Note on updating $t_{trt,i}$ in Step 3:

Let $h_{trt}(t) = \kappa_1 t^{\kappa_1 - 1} h_{0i}$, where $h_{i0} = \exp(c_0 + \alpha_1 X_i + 0.5 Z_i + \epsilon_i)$.

If $t_{ij} < t_{trt,i}$, we have $\int_{t_{ij}}^{t^*_{trt,i}} \theta h_{trt}(s) \, ds = \int_{t_{ij}}^{t_{trt,i}} h_{trt}(s) \, ds$, i.e.,

$$\int_{t_{ij}}^{t^*_{trt,i}} \theta \kappa_1 s^{\kappa_1 - 1} h_{i0} \, ds = \int_{t_{ij}}^{t_{trt,i}} \kappa_1 s^{\kappa_1 - 1} h_{i0} \, ds,$$

which implies $(t_{ij})^{\kappa_1} h_{i0} + [(t^*_{trt,i})^{\kappa_1} - (t_{ij})^{\kappa_1}] \theta h_{i0} = (t_{trt,i})^{\kappa_1} h_{i0}$. Therefore, $t^*_{trt,i} = \left[\frac{(t_{trt,i})^{\kappa_1} - (t_{ij})^{\kappa_1}}{\theta} + (t_{ij})^{\kappa_1}\right]^{1/\kappa_1}$ if $t^*_{trt,i} \leq t_{ij} + \delta$.

If $t^*_{trt,i} > t_{ij} + \delta$, then we have $\int_{t_{ij}}^{t_{ij}+\delta} \theta h_{trt}(s) \, ds + \int_{t_{ij}+\delta}^{t^*_{trt,i}} h_{trt}(s) \, ds = \int_{t_{ij}}^{t_{trt,i}} h_{trt}(s) \, ds$, i.e.,

$$\int_{t_{ij}}^{t_{ij}+\delta} \theta \kappa_1 s^{\kappa_1 - 1} h_{i0} \, ds + \int_{t_{ij}+\delta}^{t^*_{trt,i}} \kappa_1 s^{\kappa_1 - 1} h_{i0} \, ds = \int_{t_{ij}}^{t_{trt,i}} \kappa_1 s^{\kappa_1 - 1} h_{i0} \, ds,$$

which implies $(t_{ij})^{\kappa_1} h_{i0} + [(t_{ij}+\delta)^{\kappa_1} - (t_{ij})^{\kappa_1}] \theta h_{i0} + [(t^*_{trt,i})^{\kappa_1} - (t_{ij}+\delta)^{\kappa_1}] h_{i0} = (t_{trt,i})^{\kappa_1} h_{i0}$. Therefore, $t^*_{trt,i} = \left[(t_{trt,i})^{\kappa_1} + (1-\theta)\left((t_{ij}+\delta)^{\kappa_1} - (t_{ij})^{\kappa_1}\right)\right]^{1/\kappa_1}$.

Main set of simulation scenarios/parameters

Pre-match sample is set at 3000 persons. For implementation of PERR, match treated persons to control persons at 1:1 ratio on $Z_i$. The sample size for the matched PERR dataset is subject to chance and scenario parameters and usually between 2000 and 2500. Each scenario is evaluated with 1,000 replicates.

Unless otherwise specified, the parameters are $k_1 = k_2 = 1.25$, $c_0 = -8$, $\beta_0 = -7$, $\alpha_1 = \alpha_2 = 0.5$, $\tau_i \sim uniform(200, 300)$, and treatment effect $\beta = \ln(0.5)$.

i) For comparison of PERR, $PERR_{Cox}$ and $PERR_{AG}$, we include two levels of baseline event rate ($\beta_0 = -6.5, -7.5$) and three levels of variance of $w_i$ ($\sigma_\omega^2 = 0, 0.5, 1.0$). Estimation of standard error for PERR is by bootstrapping with 200 replicates, and by robust standard error for clustered data for the other two methods.

ii) For evaluation of strategy to detect event impact on treatment time, we consider two levels of $\delta$ ($\delta = 20, 30$), four levels of $\theta$ ($\theta = \frac{1}{4}, \frac{1}{2}, 2, 4$) and, for evaluation of model misspecification, two patterns of event-dependent treatment that in the first half of $\delta$ the event impact on treatment time is stronger and in the second half it is weaker, ($\theta = \frac{1}{4}, \frac{1}{2}$) and ($\theta = 4, 2$). In addition, we include a negative control scenario that has $\delta = 0$ and $\theta = 1$. In all estimation we use M = 5 and $\dot{\delta} = 10$ in equation (6).

iii) For correction of bias due to event-dependent treatment in (ii), we use the estimated $\hat{\delta}$ and $\hat{\theta}$ for making the correction for bias arising from event-dependent treatment.

Supplemental set of simulation scenarios/parameters

Further simulations are the same as the base scenarios with $\delta = 30$ and $\theta = \left(\frac{1}{4}, \frac{1}{2}, 2, 4\right)$ but the following parameters are varied:

Table S1: Using true event impact on treatment $(\delta, \theta)$, instead of their estimates $(\hat{\delta}, \hat{\theta})$, in the correction method.

Table S2: Treatment effect is $\beta = 0$.

Table S3: Treatment effect is $\beta = \ln(2)$.

Table S4: Unobserved confounder is continuous $(e^X \sim Gamma(4, 1/4))$.

Table S5: Opposite direction of confounding, $(\alpha_2 = -0.5, \beta_0 = -6.5)$.

Table S6: Stronger confounding, $(\alpha_1 = \alpha_2 = 1, c_0 = -8.5, \beta_0 = -7.5)$.

Table S7: Events have flatter hazard function $(k_2 = 1.0, \beta_0 = -5.5)$.

Table S8: Events have steeper hazard function $(k_2 = 1.5, \beta_0 = -8.5)$.

Table S9: Treatment has flatter hazard function $(k_1 = 1.0, c_0 = -6.5)$.

Table S10: Treatment has steeper hazard function $(k_1 = 1.5, c_0 = -9.5)$.

Table S11: More variable follow-up time, $\tau \sim uniform(150, 350)$.

## S3.2. Supplemental Simulation Results

Table S1. Simulation results on estimation of treatment effect by PERR$_{AG}$ with and without correction for event-dependent treatment bias, by levels of $\delta$ and $\theta$; true treatment effect = $\exp(\beta_3) = 0.5$; correction is based on the true $\delta$ and $\theta$ values instead of their estimates.

| $\delta$ | $\theta$ | Uncorrected | | | Corrected based on ($\delta,\theta$) | | |
|---|---|---|---|---|---|---|---|
| | | Mean estimate | CP (%) | RMSE | Mean estimate | CP (%) | RMSE |
| 0 [a] | 1 [a] | 0.510 | 94.8 | 0.050 | 0.509 | 94.9 | 0.049 |
| 20 | 1/4 | 0.575 | 72.7 | 0.092 | 0.484 | 93.5 | 0.051 |
| 20 | 1/2 | 0.548 | 86.0 | 0.072 | 0.504 | 94.8 | 0.050 |
| 20 | 2 | 0.468 | 90.4 | 0.053 | 0.504 | 94.8 | 0.049 |
| 20 | 4 | 0.433 | 66.1 | 0.079 | 0.504 | 93.8 | 0.054 |
| 30 | 1/4 | 0.626 | 38.6 | 0.141 | 0.474 | 91.4 | 0.058 |
| 30 | 1/2 | 0.573 | 71.4 | 0.092 | 0.505 | 88.0 | 0.066 |
| 30 | 2 | 0.454 | 82.5 | 0.063 | 0.497 | 94.3 | 0.051 |
| 30 | 4 | 0.416 | 52.0 | 0.094 | 0.493 | 94.2 | 0.053 |
| 30 [b] | 1/4, 1/2 [b] | 0.601 | 54.2 | 0.117 | 0.483 | 90.9 | 0.059 |
| 30 [c] | 4, 2 [c] | 0.435 | 70.0 | 0.077 | 0.502 | 93.9 | 0.052 |

[a] Negative control: events have no impact on treatment.
[b] $\theta = 1/4$ for 15 days followed by $\theta = 1/2$ for another 15 days.
[c] $\theta = 4$ for 15 days followed by $\theta = 2$ for another 15 days.

Tables S2. Simulation results on estimation of treatment effect with and without correction for event-dependent treatment bias, HR =1.

| $\theta$ | Uncorrected | | | Corrected based on $(\hat{\delta}, \hat{\theta})$ | | |
|---|---|---|---|---|---|---|
| | Mean HR | CP (%) | RMSE | Mean HR | CP (%) | RMSE |
| 1/4 | 1.252 | 33.5 | 0.278 | 0.947 | 88.2 | 0.116 |
| 1/2 | 1.150 | 68.4 | 0.182 | 1.013 | 86.7 | 0.125 |
| 2 | 0.910 | 79.7 | 0.120 | 1.000 | 93.5 | 0.097 |
| 4 | 0.840 | 51.5 | 0.179 | 0.994 | 93.7 | 0.101 |

Tables S3. Simulation results on estimation of treatment effect with and without correction for event-dependent treatment bias, HR =2.

| $\theta$ | Uncorrected | | | Corrected based on $(\hat{\delta}, \hat{\theta})$ | | |
|---|---|---|---|---|---|---|
| | Mean HR | CP (%) | RMSE | Mean HR | CP (%) | RMSE |
| 1/4 | 2.508 | 27.5 | 0.553 | 1.903 | 89.5 | 0.219 |
| 1/2 | 2.302 | 66.2 | 0.365 | 2.018 | 84.7 | 0.255 |
| 2 | 1.822 | 80.1 | 0.235 | 1.997 | 91.8 | 0.187 |
| 4 | 1.668 | 46.2 | 0.364 | 1.974 | 93.5 | 0.194 |

Tables S4. Simulation results on estimation of treatment effect with and without correction for event-dependent treatment bias with a continuous unobserved confounder X ($e^X \sim Gamma(4, 1/4)$), HR = 0.5.

| $\theta$ | Uncorrected | | | Corrected based on ($\hat{\delta}, \hat{\theta}$) | | |
|---|---|---|---|---|---|---|
| | Mean HR | CP (%) | RMSE | Mean HR | CP (%) | RMSE |
| 1/4 | 0.625 | 50.5 | 0.143 | 0.481 | 91.2 | 0.065 |
| 1/2 | 0.579 | 77.8 | 0.101 | 0.528 | 87.0 | 0.079 |
| 2 | 0.449 | 82.4 | 0.069 | 0.488 | 92.4 | 0.057 |
| 4 | 0.408 | 47.2 | 0.102 | 0.481 | 92.1 | 0.058 |

Tables S5. Simulation results on estimation of treatment effect with and without correction for event-dependent treatment bias with opposite direction of confounding ($\alpha_2 = -0.5, \beta_0 = -6.5$), HR = 0.5.

| $\theta$ | Uncorrected | | | Corrected based on ($\hat{\delta}, \hat{\theta}$) | | |
|---|---|---|---|---|---|---|
| | Mean HR | CP (%) | RMSE | Mean HR | CP (%) | RMSE |
| 1/4 | 0.609 | 52.0 | 0.124 | 0.460 | 86.8 | 0.064 |
| 1/2 | 0.563 | 78.8 | 0.083 | 0.497 | 86.8 | 0.065 |
| 2 | 0.439 | 72.6 | 0.074 | 0.481 | 90.8 | 0.054 |
| 4 | 0.393 | 29.5 | 0.113 | 0.469 | 88.6 | 0.059 |

Tables S6. Simulation results on estimation of treatment effect with and without correction for event-dependent treatment bias with a stronger confounding ($\alpha_1 = \alpha_2 = 1, c_0 = -8.5, \beta_0 = -7.5$), HR = 0.5.

| $\theta$ | Uncorrected | | | Corrected based on ($\hat{\delta}, \hat{\theta}$) | | |
|---|---|---|---|---|---|---|
| | Mean HR | CP (%) | RMSE | Mean HR | CP (%) | RMSE |
| 1/4 | 0.636 | 46.6 | 0.156 | 0.489 | 92.0 | 0.068 |
| 1/2 | 0.586 | 72.4 | 0.108 | 0.535 | 83.4 | 0.084 |
| 2 | 0.466 | 90.0 | 0.059 | 0.504 | 92.4 | 0.057 |
| 4 | 0.433 | 72.6 | 0.080 | 0.501 | 94.2 | 0.055 |

Tables S7. Simulation results on estimation of treatment effect with and without correction for event-dependent treatment bias when events have flatter hazard function ($\kappa_2 = 1.0, \beta_0 = -5.5$), HR = 0.5.

| $\theta$ | Uncorrected | | | Corrected based on ($\hat{\delta}, \hat{\theta}$) | | |
|---|---|---|---|---|---|---|
| | Mean HR | CP (%) | RMSE | Mean HR | CP (%) | RMSE |
| 1/4 | 0.603 | 53.2 | 0.118 | 0.471 | 90.6 | 0.056 |
| 1/2 | 0.560 | 77.9 | 0.079 | 0.495 | 89.7 | 0.057 |
| 2 | 0.469 | 88.1 | 0.053 | 0.520 | 91.5 | 0.055 |
| 4 | 0.446 | 76.3 | 0.069 | 0.539 | 87.4 | 0.069 |

Tables S8. Simulation results on estimation of treatment effect with and without correction for event-dependent treatment bias when events have steeper hazard function ($\kappa_2 = 1.5, \beta_0 = -8.5$), HR = 0.5.

| $\theta$ | Uncorrected | | | Corrected based on $(\hat{\delta}, \hat{\theta})$ | | |
|---|---|---|---|---|---|---|
| | Mean HR | CP (%) | RMSE | Mean HR | CP (%) | RMSE |
| ¼ | 0.648 | 33.5 | 0.164 | 0.480 | 91.4 | 0.061 |
| ½ | 0.588 | 67.2 | 0.107 | 0.520 | 84.7 | 0.077 |
| 2 | 0.441 | 75.1 | 0.074 | 0.479 | 90.1 | 0.057 |
| 4 | 0.393 | 35.4 | 0.114 | 0.459 | 86.4 | 0.064 |

Tables S9. Simulation results on estimation of treatment effect with and without correction for event-dependent treatment bias when treatment has flatter hazard function ($\kappa_1 = 1.0, c_0 = -6.5$), HR = 0.5.

| $\theta$ | Uncorrected | | | Corrected based on $(\hat{\delta}, \hat{\theta})$ | | |
|---|---|---|---|---|---|---|
| | Mean HR | CP (%) | RMSE | Mean HR | CP (%) | RMSE |
| ¼ | 0.642 | 30.6 | 0.156 | 0.499 | 93.5 | 0.056 |
| ½ | 0.586 | 65.3 | 0.104 | 0.527 | 86.9 | 0.071 |
| 2 | 0.448 | 78.6 | 0.068 | 0.484 | 91.5 | 0.055 |
| 4 | 0.395 | 34.9 | 0.112 | 0.458 | 84.6 | 0.066 |

Tables S10. Simulation results on estimation of treatment effect with and without correction for event-dependent treatment bias when treatment has flatter hazard function ($\kappa_1 = 1.5, c_0 = -9.5$), HR = 0.5.

| $\theta$ | Uncorrected | | | Corrected based on $(\hat{\delta}, \hat{\theta})$ | | |
|---|---|---|---|---|---|---|
| | Mean HR | CP (%) | RMSE | Mean HR | CP (%) | RMSE |
| 1/4 | 0.615 | 49.5 | 0.130 | 0.455 | 85.0 | 0.069 |
| 1/2 | 0.568 | 76.9 | 0.089 | 0.470 | 86.8 | 0.068 |
| 2 | 0.462 | 86.2 | 0.058 | 0.513 | 92.6 | 0.054 |
| 4 | 0.433 | 65.1 | 0.080 | 0.525 | 90.0 | 0.063 |

Tables S11. Simulation results on estimation of treatment effect with and without correction for event-dependent treatment bias with a more variable follow-up time $\tau \sim$ uniform(150, 300), HR = 0.5.

| $\theta$ | Uncorrected | | | Corrected based on $(\hat{\delta}, \hat{\theta})$ | | |
|---|---|---|---|---|---|---|
| | Mean HR | CP (%) | RMSE | Mean HR | CP (%) | RMSE |
| 1/4 | 0.627 | 40.5 | 0.142 | 0.462 | 85.3 | 0.065 |
| 1/2 | 0.578 | 70.6 | 0.096 | 0.503 | 84.4 | 0.069 |
| 2 | 0.465 | 86.2 | 0.057 | 0.510 | 92.2 | 0.055 |
| 4 | 0.423 | 60.7 | 0.087 | 0.504 | 94.6 | 0.053 |